\documentclass[12pt]{article}

\input epsf

\newcommand{\beq}{\begin{eqnarray}}
\newcommand{\eeq}{\end{eqnarray}}

\newcommand{\bi}{\bibitem}

\newcommand{\mC}{{\mathbf C}}
\newcommand{\mR}{{\mathbf R}}
\newcommand{\mZ}{{\mathbf Z}}
\newcommand{\mA}{{\mathbf A}}
\newcommand{\mB}{{\mathbf B}}
\newcommand{\mV}{{\mathbf V}}
\newcommand{\mU}{{\mathbf U}}
\newcommand{\bphi}{{\bf \Phi}}
\def\be{\begin{equation}}
\def\ee{\end{equation}}
\def\bea{\begin{eqnarray}}
\def\eea{\end{eqnarray}}
\def\trsm{{\rm\ Tr\,}}
\def\tr{{\bf\ Tr \,}}
\def\S{{\bf S}}
\newcommand{\newsection}[1]{
\vspace{10mm}
\pagebreak[3]
\addtocounter{section}{1}
\setcounter{subsection}{0}
\noindent
{\large\bf \thesection. #1}
\nopagebreak
\medskip
\nopagebreak}
\newcommand{\newsubsection}[1]{
\vspace{5mm}
\pagebreak[3]
\addtocounter{subsection}{1}
\addcontentsline{toc}{subsection}{\protect
\numberline{\arabic{section}.\arabic{subsection}}{#1}}
\noindent{\em 
\thesubsection. #1}
\nopagebreak
\vspace{2mm}
\nopagebreak}
\newcommand{\figuren}[3]{\addtocounter{figure}{1}
\begin{figure}[htb]\begin{center}
\leavevmode\hbox{\epsfxsize=#2 \epsffile{#1.eps}}\\[3mm] \bigskip
\parbox{14.5cm}{\small \bf Fig.\thefigure.\ \it  #3}\vspace{-5mm}
\end{center}\end{figure}}

\title{{\bf Strings from Quivers,}\\[2mm]
{\bf Membranes from Moose}}

\author{Sunil Mukhi$^{a,b,}$\footnote{On 
sabbatical leave from 
Tata Institute of Fundamental 
Research, 10/2001--9/2002.}~, Mukund Rangamani$^c$ and Erik
Verlinde$^c$\\ \\
\small \sl $^a$School of Natural Sciences, Institute for Advanced 
Study,\\[-1.5mm]
\small \sl Princeton, NJ 08540, USA\\ 
\small\sl $^b$Tata Institute of Fundamental Research,\\[-1.5mm]
\small\sl Mumbai 400 005, India\\ 
\small \sl $^c$Department of Physics, Princeton University\\[-1.5mm]
\small \sl Princeton, NJ 08544, USA\\}

\begin{document}
\setlength{\baselineskip}{16pt}
\begin{titlepage}
\maketitle
\begin{picture}(0,0)(0,0)
\put(325,330){hep-th/0204147}
\put(325,315){PUPT--2033}
\put(325,300){TIFR/TH/02-12}
\end{picture}
\begin{abstract}

We consider ${\cal N}=2$ moose/quiver gauge theories corresponding to
$N_1$ D3-branes at a $\mC^2/\mZ_{N_2}$ singularity in the ``large
moose'' limit where $N_1$ and $N_2$ are scaled to infinity together.
In the dual holographic description, this scaling gives rise to
a maximally supersymmetric pp-wave background with a {\em compact}
light-cone direction. We identify the gauge theory operators that
describe the Discrete Light-Cone Quantization (DLCQ) of the string in
this background.  For each discrete light-cone momentum and winding
sector there is a separate ground state and Fock space. The large
moose/quiver diagram provides a useful graphical representation of the string
and its excitations. This representation has a natural explanation in
a T-dual language. The dual theory is 
a non-relativistic type IIA string wound around the T-dual
direction, and bound by a quadratic Newtonian potential. We end with
some comments on D-string/D-particle states, a possible lift to
M-theory and the relation to deconstruction.

\end{abstract}
\thispagestyle{empty}
\setcounter{page}{0}
\end{titlepage}

\renewcommand{\baselinestretch}{1.2}  

\newsection{Introduction}

It is believed that gauge theories encode the dynamics of
strings. Attempts to physically realize this idea led to the large-N
expansion of gauge theory \cite{tHooft}. In this context, string
theories have indeed been seen to emerge from gauge theories, most
notably in Matrix Theory \cite{BFSS,IKKT} and in the AdS/CFT
correspondence \cite{Juan}. 

Matrix theory gives, in principle, a precise definition for
non-perturbative M-theory using large-N supersymmetric quantum
mechanics. Toroidal compactifications of M-theory are then described
by large-N gauge theories in various dimensions. The construction
of strings in terms of gauge fields can be made rather explicit in
some cases. The best example is Matrix String
Theory \cite{Motl,BS,DVV}. Here, perturbative string states and also
D-branes in type IIA string theory can be identified in terms of
operators of a two-dimensional ${\cal N}=8$ supersymmetric gauge
theory.

The AdS/CFT correspondence provided a very explicit map between gauge
theory operators and string theory states in the supergravity
limit \cite{GKP,EW}. Although the proposal can be implicitly extended
to the full string theory, it proves difficult to construct strings
explicitly out of gauge fields in this framework. Recently, new
insight was obtained by considering a special limit of the $AdS$
background of type IIB string theory, in which it is possible to
construct string states from gauge theory operators \cite{BMN}. This is
achieved by taking a Penrose limit \cite{Penrose} of the background,
which provides a maximally supersymmetric solution of type IIB string
theory \cite{Blau}, and relating this to a special subset of operators
in ${\cal N}=4$ supersymmetric gauge theory. 

Penrose limits of spacetimes generally lead to pp-wave backgrounds,
which are exact backgrounds for string propagation \cite{AK,HS} and are
exactly solvable in the light-cone gauge \cite{HStwo,Metsaev}. In
light-cone quantization, it is often useful to compactify a null
direction. This leads to the Discrete Light-Cone Quantization (DLCQ),
which provides a convenient regulator for string
theory \cite{Thorn}. In this picture, one has interacting strings
carrying quantized units of the light-cone momentum, with the minimal
momentum being carried by a ``string bit''. The idea of DLCQ is most
natural in the context of Matrix Theory \cite{Lenny}, where it has been
argued that each sector of the DLCQ of M-theory is exactly described
by a $U(N)$ matrix model. Although a compact null direction may look
somewhat strange, it is best thought of as the limit \cite{Nati} of a
spacelike compactification.

Our aim is to study type IIB string theory in a DLCQ pp-wave
background. The Penrose limits that have been considered so
far in the AdS/CFT context always lead to a pp-wave with a
non-compact null direction. We will show that there is a novel scaling
limit of a particular $AdS$ background, which precisely gives rise to
a DLCQ pp-wave. The radius of the null direction is a finite
controllable parameter of this background. The corresponding gauge
theory is an ${\cal N}=2$ superconformal ``moose'' or ``quiver''
theory \cite{DM}.

A number of fascinating aspects of the gauge theory/pp-wave
correspondence will unfold as we explore this model. As in previous
examples following the original ideas of Ref.\cite{BMN}, we find
gauge theory operators which can be identified to the string ground
state, zero-mode oscillators and excited oscillator states. However,
there are remarkable differences, and a much richer structure appears
due to the DLCQ nature of the string background. We find a gauge
theory operator that describes a string ground state for every value
of the (positive) quantized light-cone momentum $k$. We will also
construct operators that describe winding states of the string over
the null direction. The gauge theory operators automatically possess
all the properties required of DLCQ string states.

Our construction turns out to have a tantalizing property: the
operators that describe DLCQ momentum states have the structure of a
``string'' of operators that winds around the moose/quiver
diagram. This is suggestive of a string winding state. Conversely, the
gauge theory operators that describe winding of the DLCQ string look
very much like momentum states.  This suggests that we should consider
T-dualization of the DLCQ pp-wave to find the most direct
correspondence with the gauge theory. We perform the requisite
T-duality and find a theory of non-relativistic
strings \cite{klebmal,gomisnr,danielsson} bound in a harmonic-oscillator
potential. The strings are wound around a compact spatial direction
which can be identified with the ``theory space'' direction of the
quiver. On lifting to M-theory we find non-relativistic membranes
winding around a 2-torus and bound in a potential. Both these
observations suggest that the moose/quiver theory, in our scaling
limit, provides a precise realization of the concept of
``deconstruction'' put forward in 
Refs.\cite{Arkani,Pokorski,ACKKM,Rothstein}.

This paper is organized as follows. In Section 2, we describe the
moose/quiver gauge theory and its $AdS$ dual in some detail, and
exhibit the scaling limit that we consider. In Section 3, we
discuss the DLCQ pp-wave background and show how it arises in our
scaling limit. We also discuss general properties of string propagation
in this background. Section 4 is devoted to the identification of
spacetime quantum numbers with charges of gauge-theory operators. In
Section 5 we explicitly construct gauge theory operators to describe
various string excitations in the DLCQ pp-wave. Section 6 deals with
T-duality of the null direction to obtain a non-relativistic string,
while Section 7 gives a physical interpretation of the relationship
between quivers and strings. In Section 8 we make some conjectural
remarks about how D-strings and other non-perturbative objects may be
described in our formalism, and exhibit the lift of the IIA
non-relativistic string to M-theory. Finally, we relate our work to the
deconstruction idea, and close with some comments.

\newsection{The Large Moose/Quiver Theory and its Holographic Dual}

It has been known for some time that one can get four-dimensional
conformal field theories by placing D3-branes at
orbifolds \cite{DM}. These theories admit an $AdS$ dual where the
compact 5-manifold is an orbifold of $\S^5$ \cite{KS}. The specific
case we will consider here is obtained by starting with $N_1$
D3-branes transverse to the 6-dimensional space $\mC^3/\mZ_{N_2}$. In
the covering space there are $N_1N_2$ D3-branes, which define a
``parent'' ${\cal N}=4$ super-Yang-Mills theory, of which the orbifold
theory is a projection.

The orbifold group $\mZ_{N_2}$ acts on $\mC^3$ by:
\be
(z_1,z_2,z_3)\rightarrow (z_1,\omega z_2, \omega^{-1}z_3), 
\qquad \omega= e^{{2\pi i\over N_2}}
\ .
\label{aledef}
\ee
The theory on the brane world-volume is a ${\cal N}=2$ superconformal
field theory in four dimensions, with the R-symmetry group
$U(1)_R\times SU(2)_R$. The gauge group
is 
\be
SU(N_1)^{(1)}\times SU(N_1)^{(2)} \times\cdots SU(N_1)^{(N_2)}.
\ee 
The fields in the vector multiplet for each
factor of the gauge group are denoted $(A_{\mu I}, \Phi_I, \psi_{aI})$
with $I$ labelling the gauge group, $I = 1, \cdots N_2$, 
and $a=1,2$. In addition, there are hypermultiplets
$(A_I, B_I,
\chi_{aI})$, where the $A_I$ are bi-fundamentals in the $(N_1,{\bar
N}_1)$ of $SU(N_1)^{(I)}\times SU(N_1)^{(I+1)}$ and the $B_I$ are
bi-fundamentals in the complex conjugate representation $({\bar N_1},
N_1)$. 
The matter content of the gauge theory can be succintly summarised 
in the form of a quiver/moose diagram, see Fig.1. Here we use the 
${\cal N }=1 $ language to describe the ${\cal N } = 2 $ gauge theory, which 
will be of use later. The fields $\Phi_I$, $A_I$ and $B_I$ can be 
identified with the $z_1, z_2$ and $z_3$ directions of the $\mC^3$.
\figuren{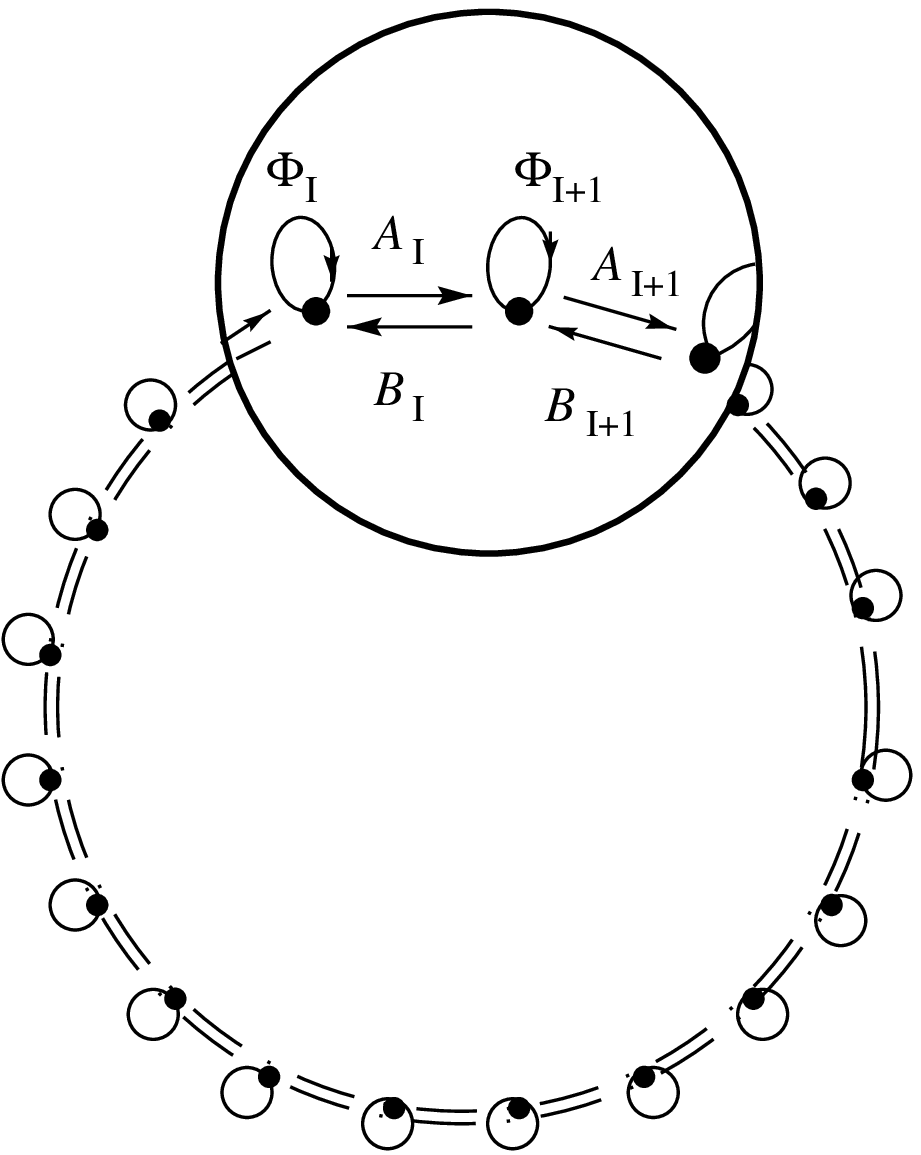}{7cm}{In the large $N_2$ limit the moose/quiver 
diagram of the gauge theory contains an large number of nodes. The two
lines connecting each pair of nodes correspond to the bifundamental
fields $A_I$ and $B_I$ and the line going back to the same node
represents the adjoint scalars $\Phi_I$.}

The holographic dual of the quiver theory in question is type IIB string 
theory on $AdS_5\times
\S^5/\mZ_{N_2}$. The action of $\mZ_{N_2}$ is obtained by thinking of
the 5-sphere as embedded in $\mR^6\sim \mC^3$ where the action is 
as prescribed in (\ref{aledef}).
This leaves a fixed circle along an equator of $\S^5$. 
The $AdS_5$ space has a radius given by
\be
R^2  = \sqrt{4\pi g^B_s \alpha'^2 N_1 N_2}
\ ,
\label{adsradius}
\ee
where $g^B_s$ is the type IIB string coupling. There are also $N_1N_2$
units of 5-form flux through the $AdS_5$.

Because of the orbifold action, the volume of $\S^5/Z_{N_2}$ is
reduced by a factor $N_2$ compared to that of the covering space
$\S^5$, with the latter having the same radius as that of $AdS_5$
given in Eq.(\ref{adsradius}). Similarly, there are $N_1$ units of
5-form flux through the $\S^5/Z_{N_2}$ factor, which descend from
$N_1N_2$ units of flux in the covering space.

It is also worth noting that the coupling constant in each of the 
gauge group factors is given in terms of the Type IIB 
coupling constant as 
\be
(g_{YM})_I^2 = { 4 \pi g_s^B  N_2}
\ .
\ee
This means that the 't Hooft coupling relevant for each factor is
\be
\lambda = (g_{YM})_I^2\, N_1 = 4\pi g^B_s N_1 N_2
\ .
\ee
This is the same as the 't Hooft coupling on the original $N_1N_2$
D3-branes before orbifolding, for which the Yang-Mills coupling was
equal to $4\pi g^B_s$.

In the following, we will consider a scaling limit when both $N_1$ and
$N_2$ become large, with the ratio $N_1\over N_2$ held fixed. In this
limit the 't Hooft coupling $\lambda$ diverges. As argued in
Ref.\cite{BMN} in the context of $SU(N)$ super-Yang-Mills theory with
${\cal N}=4$ supersymmetry, the relevant quantity that needs to be
kept finite, is $g^B_s N\over J^2$ where $J$ is a $U(1)$ charge and
the relevant states have very large $J$. In our case, we will see that
the role of $J$ is played by $N_2$, while $N$ is replaced by
$N_1N_2$. So the quantity that should be kept finite for us is $g^B_s
N_1\over N_2$. This is achieved precisely by scaling $N_1$ and $N_2$
to infinity and keeping $g^B_s$ small but finite.

\newsection{The DLCQ pp-Wave}

In this section we start from the holographic description of the 
quiver gauge theory under consideration. We will take a 
limit of the dual spacetime, which is known as the Penrose limit 
 \cite{Penrose} ({\it cf.} \cite{Gueven}, for generalization to 
supergravity). The essential idea is to consider  a 
null geodesic and look at the spacetime in the neighbourhood of the 
geodesic. It was first demonstrated by Penrose that this is a 
sensible limit to consider in any geometry and the result is always 
what is known as a plane-parallel wave or pp-wave for short. 

\newsubsection{The Penrose Limit of $AdS_5\times \S^5/\mZ_{N_2}$}

As mentioned above, to obtain the Penrose limit of any gravitational 
background, one has to
focus on a light-like geodesic. In the particular 
case of $AdS_5 \times \S^5/\mZ_{N_2}$, we choose a null geodesic 
which is based at the origin
of $AdS_5$ and carries some angular momentum along the compact directions. 
Because of the singular nature of the compact manifold,
the result depends on whether one takes this trajectory to lie along
the singular locus or not. The former choice results in a pp-wave
background that has the $\mZ_{N_2}$ ALE singularity as part of its
transverse space \cite{IKM}-\cite{FK}. The latter
choice, which was briefly discussed in Ref.\cite{Ali}, results instead
in the maximally supersymmetric pp-wave background. In this paper we
focus on the latter case as it is relevant for the scaling limit
in which we are interested.

Let us write the metric of $AdS_5\times \S^5/\mZ_{N_2}$ as:
\bea
ds^2 &=& R^2\Bigg[-\cosh^2\rho \, dt^2 + d\rho^2 + \sinh^2\rho\,
d\Omega_3^2 + \nonumber\\
&& d\alpha^2 + \sin^2\alpha\, d\theta^2 + {\cos^2\alpha}\, 
\left(d\gamma^2 +
\cos^2\gamma\, d\chi^2+\sin^2\gamma\, d\phi^2\right)\Bigg]
\ ,
\label{adsmet}
\eea
where the first line is the $AdS_5$ metric in global coordinates,
while the remaining terms describe the metric for an $\S^5$ embedded in
a 6-dimensional space containing a $\mZ_{N_2}$ ALE singularity. The
relationship with the complex $z_1$ coordinates and the angles is: 
\be
z_1=R \sin\alpha\, e^{i\theta}, \quad z_2=R\cos\alpha \cos\gamma\,
e^{i\chi}, \quad z_3=R\cos\alpha \sin\gamma\, e^{i\phi}. 
\ee 
In this
parametrization the orbifold is obtained by demanding that the angles
$\chi$ and $\phi$ are periodic modulo $2\pi$ but in addition have a
combined periodicity under 
\be
\chi\to \chi+{2\pi \over N_2},\qquad \phi\to \phi-{2\pi\over N_2} \ .
\label{periodic}
\ee 
Note that with this choice there are no explicit factors of
${N_2}$ occurring in the metric.

To take the pp-wave limit, we now define new coordinates $r,w,y$
by
\bea
&r=\rho R,~~w= \alpha R, & y=\gamma R
\ .
\label{pplimit}
\eea
and introduce the light-cone coordinates
\bea
& x^+ = \frac12\left(t+\chi\right), & 
x^- = {R^2\over 2}\left(t-\chi\right)
\label{lcone}
\ .
\eea 
Making the substitutions the metric (\ref{adsmet}) becomes
\bea
ds^2 &=& R^2\Bigg[ - \cosh^2 {r \over R} \,
\left(dx^+ + {1\over R^2} dx^- \right)^2 + {dr^2 \over R^2} + 
\sinh^2 {r \over R}\,
d\Omega_3^2 + {dw^2 \over R^2} +\nonumber\\
 && \hspace{-5mm}  \sin^2{w \over R}\, d\theta^2 + 
\cos^2{w \over R} \, 
\left({dy^2 \over R^2}+
\cos^2{y \over R}\, \left(dx^+ - {1\over R^2} dx^- \right)^2+
\sin^2{y \over R}\, d\phi^2\right)\Bigg].
\label{newadsmet}
\eea
In the limit $R\rightarrow \infty$ the metric reduces to
 \cite{Ali}
\be
ds^2=-4dx^+ dx^- - (r^2+w^2+y^2) \,{dx^+}^2 + dr^2+r^2d\Omega_3^2 +
dw^2+w^2d\theta^2
+dy^2+y^2d\phi^2 
\ee
This is just the universal pp-wave background which has been found in
many other cases.  It can be written in the standard form
\be
ds^2 = -4dx^+ dx^- - \sum_{i=1}^8 (x^i)^2 \,{dx^+}^2 + \sum_{i=1}^8
{dx^i}^2 
\ , 
\label{ppwaveg}
\ee
where we introduced the eight transversal coordinates $x^i$. 
There is also a Ramond-Ramond flux in the geometry (\ref{ppwaveg}):
\be
F_{+1234} = F_{+5678} = const
\ .
\label{rrflx}
\ee

Although our model gives rise to the standard pp-wave metric in the
Penrose limit, there is actually an important difference: the lightlike
direction $x^-$ is {\em compact}. {}From
Eq.(\ref{periodic}), the $2\pi\over N_2 $ periodicity of the angle $\chi$
translates into the following periodicity condition on the light-cone
coordinates 
\bea
&& x^+\rightarrow x^+ + {\pi\over N_2}\nonumber\\
&& x^- \rightarrow x^- + {\pi R^2\over N_2}
\ .
\eea
This combined shift in $x^+$ and $x^-$ has to be accompanied by a
simultaneous shift in $\phi\to \phi-{2\pi\over N_2}$.  Under a scaling
$N_1 \sim N_2$, we have $R^2\sim N_2$. Now if $N_2 \rightarrow \infty$
we see that $x^-$ is
periodic in the limit, with period\footnote{Note that $x^-$ as defined
has dimensions of $(\rm length)^2$.}:
\be
{\pi R^2\over N_2} = 2\pi R_-,\quad R_-= \left(\pi
g^B_s\,{N_1\over N_2}\right)^\frac12\alpha'  
\ . 
\label{rminus}
\ee
After taking the limit, there are no longer accompanying shifts in
$x^+$ and $\phi$.  So the periodic direction has become pure
lightlike. As a consequence the corresponding light-cone momentum
$2p^+$ is quantized in units of $1\over R_-$. The way the compact 
null direction arises from a limit of a spacelike circle is exactly as 
discussed in \cite{Nati} in the context of Matrix theory.

In other words, the Penrose limit of $AdS_5\times \S^5/\mZ_{N_2}$ with
 $N_2\to \infty$  leads to a Discrete Light-Cone Quantization (DLCQ)
of the string on a pp-wave background, in which the null
direction $x^-$ is periodic. We now turn to a discussion of string 
propagation in such a spacetime.

\newsubsection{String Propagation in DLCQ pp-wave} 

String propagation in pp-wave backgrounds is a subject which has been 
explored extensively in the literature.
A very interesting 
fact about these backgrounds is that they are solutions to the world-sheet 
beta function equations (in covariant quantization) to 
all orders \cite{AK, HS}, thereby 
being exact string backgrounds. The pp-wave 
backgrounds admit a covariantly constant null Killing vector, implying that 
one can always choose to quantize the sigma model in   
light-cone gauge \cite{HStwo}. One can extend the analysis to strings on 
Ramond-Ramond backgrounds following \cite{Metsaev}.

Let us begin by writing down the sigma model action for the 
pp-wave geometry (\ref{ppwaveg}): 
\be
S = - {1 \over 4 \pi \alpha'} \int d\sigma \, d\tau \; 
\left(  -4 \partial_a x^+ \partial^a x^- + 
\partial_a x^i \partial^a x^i -  \sum_{i=1}^{8}(x^{i})^2 \, 
\partial_a x^+ \partial^a x^+
\right) 
\ .
\label{sigmamod}
\ee
The worldsheet equations of motion resulting from this are
\bea
\partial_a\partial^a x^+ &=& 0 
\nonumber \\
\partial_a \partial^a x^i  -  x^i (\partial_a x^+ \partial^a x^+) 
&=& 0 
\ .
\eea
In the light-cone gauge the first equation is solved by $x^+=\tau$, where 
at the same time one takes $\sigma$ to range from 0 to $2 \pi\alpha' p^+$.
In addition, in order to maintain reparametrization
invariance, one requires that the world-sheet
stress tensor vanishes. This can be expressed as:
\bea
2 \partial_\tau x^- &=& \frac12\,\left(
(\partial_\tau x^i)^2 + (\partial_\sigma x^i)^2 + 
(x^i)^2\right) \nonumber \\
2 \partial_\sigma x^- & =& \,\partial_\tau x^i \; \partial_\sigma x^i
\label{stcons}
\eea
{}From these equations we can easily derive the light-cone Hamiltonian
for the string and also derive the momentum constraints. Note that the
presence of the mass term $\sum_{i =1}^8 \, (x^i)^2$, for the
transverse scalars inhibits the separation of the scalars into left
and right movers. So far we have ignored the world-sheet fermions in
the discussion. Just like in the standard pp-wave background they are
given by massive Dirac fermions in the light-cone gauge. They get
their mass from the Ramond-Ramond background flux, supporting the
pp-wave geometry (\ref{rrflx}).
 
The solution to the world-sheet theory proceeds in the usual 
way by normal mode expansion and introduction of oscillators. In particular, 
mode expansion of the transverse coordinates $x^i$ is
\bea
x^i(\sigma, \tau)  &=& \sum_{n = -\infty}^{\infty} \; 
a^i_n \, {1 \over \sqrt{\omega_n}} \,e^{i {n \over p^+ \alpha'}
 \sigma + i \omega_n \tau } + 
{\rm h.c.}
\nonumber \\
\omega_n &=& \sqrt{1 + {n^2 \over (p^+)^2\alpha'^2}}.
\eea

One very important fact is that with the compact $x^+$ direction, the
light-cone momentum $p^+$ is quantized. This means that we have a
positive integer $k$ labelling our states, with 
\be
2 p^+ = { k \over R_-}
\ .
\ee
As is well-known, the theory then splits into sectors,
labelled by a discrete number parametrising the light-cone momentum.

The Hamiltonian and total momentum of the world-sheet theory are given as 
\bea
H &=& \sum_{n = -\infty}^{\infty} \,N_n \sqrt{1
+  {4n^2  R_-^2\over k^2 \alpha'^2}} \nonumber \\
P &=& \sum_{n= -\infty}^{\infty} \, n N_n
\ ,
\eea
where $N_n$ is the total occupation number of the Fourier mode labeled by 
$n$. These are related to the usual Virasoro generators as 
$L_0 = \frac12 ( H_{l.c} + P )$ and $\bar{L}_0 = \frac12 (H_{l.c} - P)$.

Since we are dealing with strings, we should also expect to find
states with non-zero winding number $m$. These arise as follows. If we
expand $x^-$ in a mode expansion we will get oscillators $a_n^-$,
which can be solved in terms of the transverse scalars.  In addition
we can have a zero-mode piece $ m \sigma R_-$, the usual winding term,
since $x^-$ is compact. Note that $m$ can take any integral value,
positive or negative. So we can label our string states as $\mid k,
m\rangle$. These will be our string ground states in the sector with
DLCQ momentum $k$ and winding number $m$. We can further act on these
states by the transverse oscillators to build other string states. A
general string state can therefore be denoted as
\be
\prod_{j = 1}^{M} \, a_{n_j}^{\dagger} \mid k,m\rangle
\ , 
\label{stgst}
\ee
where for convenience of notation we dropped the
superscript $i$, denoting the particular transverse coordinate on the
oscillators.  

The world-sheet reparametrization invariance gives a
constraint on the action of the oscillators for the 
states written in Eq.(\ref{stgst}). This arises from the
second equation in (\ref{stcons}) and implies
\be
 \sum_{j =1 }^{M} \, n_j = k\,m.
\label{momcons}
\ee
In the next section, we will turn our attention to the construction of 
operators in the gauge theory which will describe the string states.

\newsection{Identification of Charges and Light-cone Momenta}

As has been noted in other cases  \cite{IKM,Gomis,Zayas}, a maximally
supersymmetric pp-wave limit implies that the gauge theory has a
sector which is maximally supersymmetric.  A key ingredient in
identifying this sector of the moose/quiver gauge theory is the
interpretation of the light-cone momenta $p^+$ and $p^-$ in terms of
the global symmetries of the gauge theory.  

The $R$-symmetry group of the ${\cal N}=2$ quiver gauge theory is
$SU(2)_R \times U(1)_R$. Now, recall that $\Phi$ is associated with the
$z_1$ direction of the $\mC^3$, while $A$ and $B$ are related to $z_2$ and
$z_3$ respectively. The $U(1)_R$ factor corresponds to the
transformation $z_1\to e^{i\xi}z_1$, and therefore acts as
phase rotations on the $\Phi$ scalars. The
$A$ and $B$ fields have charge zero under this $U(1)_R$.
The
$SU(2)_R$ symmetry acts on the $A$ and $B$ fields and their complex
conjugates. Indeed these fields form a ${\cal N}=2$ hypermultiplet,
which is known to have a quaternionic structure. In fact, $(A, \bar{B})$ 
as well as $(\bar{A}, B)$ form doublets under $SU(2)_R$. Hence,
one of the generators
of this $SU(2)_R$ acts on $A$ and $B$ as phase rotations. We will denote
this generator by $J'$.  In addition there is a $U(1)$ symmetry that
is not an $R$-symmetry. This is the $U(1)$ symmetry that rotates $A$
and $B$ in opposite directions, and corresponds to $z_2\to e^{i\xi}
z_2, z_3\to e^{-i\xi}z_3$. But, because of the orbifold identification in the 
$(z_2, z_3)$ directions (\ref{aledef}), having $\xi = {2 \pi \over N_2}$ 
brings us back to the same point. Its generator $J$ together with the
$U(1) \subset SU(2)_R$ generator $J'$ will 
appear in the definition of the light-cone momentum.

In terms of the coordinates on the $\S^5/\mZ_{N_2}$ the generators $J$
and $J'$ are given by 
\be 
J=-{i\over
2N_2}\left(\partial_\chi-\partial_\phi\right),\qquad J'=-{i\over 2}
\left(\partial_\chi+\partial_\phi\right).
\ee
This leads to the following identifications for the light-cone 
momenta
\bea
2p^- =& {i} (\partial_t +\partial_\chi) ~=& \Delta
- N_2 J -J'  \nonumber\\[2mm]
2 p^+ =& \displaystyle 
i {(\partial_t - \partial_\chi)\over
R^2}  ~=& {\Delta
+ N_2 J +J' \over R^2}
\ .
\label{pplusminus}
\eea
In our conventions, the light-cone Hamiltonian is $H=2p^-$.  

As explained in Ref.\cite{BMN}, to relate gauge theory operators to
string states, we need to look for operators that have both $p^-$ and
$p^+$ finite. Since $R\rightarrow \infty$, this means $\Delta$ and
$N_2J+J'$ must both be large, while their difference remains
finite. Physical gauge invariant operators should have half-integral
values for $J$ and $J'$. This implies that $N_2J$ automatically
becomes large when $N_2\to \infty$ even when $J$ is kept
fixed. We will see that $J'$ also grows like $N_2$. The scaling dimension 
$\Delta$ also becomes large automatically, because we have the BPS
bound $\Delta\geq N_2J+J'$.  In fact, if we keep $\Delta-N_2J-J'$
fixed, then both quantities precisely grow in the right way for the
pp-wave limit, provided we take $N_1$ and $N_2$ simultaneously to
infinity with the ratio $N_1/N_2$ fixed.  In this double scaling limit
we have $R^2\sim \sqrt{N_1N_2}\sim N_2$, and so indeed $p^+$ and $p^-$
stay both finite.  We would like to emphasize again that, in contrast
with the other pp-wave backgrounds considered so far, there is no need
in our case to send $J$ to infinity. In fact, in order to reproduce
the DLCQ spectrum we have to keep $J$ finite, since it will give the
discrete value of the light-cone momentum. We have $J = {\frac12 k}$.

We now discuss the $(\Delta, J, J')$ eigenvalues of the various local
operators in the gauge theory, and construct the string ground state
and oscillators in terms of these. Because of ${\cal N}=2$
supersymmetry, all the fundamental bosonic fields have exact conformal
dimension 1, the same as their free field value, while the fermions
similarly have dimension ${3\over 2}$.

The charges are obtained as follows. The $A_I$ and $B_I$ fields that
make up the hypermultiplets have fractional charge under $J$. The
reason is that $e^{4\pi i J}$ precisely generates the orbifold
transformation $z_2\to\omega z_2$, $z_3\to\omega^{-1} z_3$.  The $A$
and $B$ fields transform accordingly, and hence have charge ${1\over
2N_2}$ and $-{1\over 2N_2}$ respectively.  The operator $J'$ generates
a $U(1)$ symmetry contained in the $SU(2)_R$ factor of the R-symmetry
group $U(1)_R\times SU(2)_R$. Under this $U(1) \subset SU(2)_R$, 
the fields $\Phi_I$
are neutral since they correspond to translations of the original
$N_1 N_2$ D3-branes along the transverse $\mR^2$ that is unaffected by
the orbifold group. On the other hand, the scalars $A_I, B_I$ in the
hypermultiplets both have charge $\frac12$ under $J'$. Complex
conjugation and supersymmetry give us the remaining charge
assignments, for the the fermions and all the conjugate fields.

\begin{table}[t]
\begin{center}
\begin{tabular}[t]{c|c|c|c|c|p{1cm} c|c|c|c|c|}
~ & $\Delta$ & $N_2 J$ & $J'$ & $H$ & & & $\Delta$ & $N_2 J$ & $J'$ 
& $H$ \\[1mm]
\cline{1-5}\cline{7-11} &&&&&&&&&& \\[-4mm]
$A_I$ & 1 & ${1\over 2}$ &  ${1\over 2}$ & 0 
&& ${\overline A_I}$ & 1 & $-{1\over 2}$ &  
$-{1\over2}$ &
2 \\[1mm]
\cline{1-5}\cline{7-11} &&&&&&&&&& \\[-4mm]
$B_I$ & 1 & $-{1\over 2}$ &  ${1\over 2}$ &  1 
&& ${\overline B_I}$ & 1 & ${1\over 2}$ &  
$-{1\over2}$ & 
1 \\[1mm]
\cline{1-5}\cline{7-11} &&&&&&&&&& \\[-4mm]
$\Phi_I$ & 1 & 0 &  0  &  1
&& ${\overline \Phi_I}$ & 1 & 0 &  0  &  
1 \\[1mm]
\cline{1-5}\cline{7-11} &&&&&&&&&& \\[-4mm]
${\chi_{A}}_I$ & ${3\over 2}$ & ${1\over 2}$ &  0 &  1 
&& ${\overline{\chi}_A}_I$ & ${3\over 2}$ & $-{1\over 2}$ &  
0 & 2 \\[1mm]
\cline{1-5}\cline{7-11} &&&&&&&&&& \\[-4mm]
${\chi_{B}}_I$ & ${3\over 2}$ & $-{1\over 2}$ &  0 &  2 
&& ${\overline{\chi}_B}_I$ & ${3\over 2}$ & ${1\over 2}$ &  0
& 1 \\[1mm]
\cline{1-5}\cline{7-11} &&&&&&&&&& \\[-4mm]
${\psi_\Phi}_I$ & ${3\over 2}$ & 0 &  $-{1\over 2}$ & 2
&& ${\overline {\psi}_\Phi}_I$ & ${3\over 2}$ & 0 &  
${1\over 2}$ & 1 \\[1mm]
\cline{1-5}\cline{7-11} &&&&&&&&&& \\[-4mm]
$\psi_I$ & ${3\over 2}$ & 0 & $-{1\over2}$ & 2
&& $\overline\psi_I$ & ${3\over 2}$ & 0 & 
${1\over 2}$ & 1 \\[1mm]
\cline{1-5}\cline{7-11}
\multicolumn{5}{c}{\vtop{\smallskip
\hbox{\small Table 1: Dimensions and charges for}
\smallskip
\hbox{~~~~~~~~~~~\small chiral fields and gauginos}}} 
& \multicolumn{6}{c}{~~~~~~~~~\vtop{\smallskip\hbox{\small 
Table 2: Dimensions and charges 
for}\smallskip\hbox{~~~~~~~~~~~\small complex conjugate fields}}} \\
\end{tabular}\\
\end{center}
\end{table}

The dimension and charge assignments, along with the $H=2p^-$ values,
are summarized in Tables 1 and 2.
In Table 1, $A_I, B_I$ refer to the scalar components of the ${\cal N}=1$ 
chiral superfields that form the ${\cal N}=2$ 
hypermultiplets. $\chi_{A_I},\chi_{B_I}$ are their fermionic
partners. $\Phi_I$ are the complex scalars in the vector multiplet,
while ${\psi_\Phi}_I$ are their fermionic partners. Finally, $\psi_I$
are the gauginos in the theory. Table 2 lists the complex conjugate
fields.

\newsection{String States from Gauge Theory Operators}

\noindent
The duality between type IIB string theory and the quiver gauge theory
implies that all string states must have corresponding states in the
gauge theory. As emphasized in Ref.\cite{BMN} one can use radial
quantization to map the gauge theory states on $\S^3\times \mR$ to
local operators. Therefore, string states are holographically dual to
operators in the gauge theory.  In this section we will identify these
operators in the large moose/quiver theory and discuss how they match
the string spectrum in the DLCQ pp-wave background.  The construction
will be similar in spirit to the BMN-operators\footnote{ The acronym BMN 
refers to the authors of Ref.\cite{BMN}, {\it viz.}, 
Berenstein, Maldacena and Nastase.} \cite{BMN} in the
${\cal N}=4$ case, but there will be important differences in the
details. One of the main differences is that in our case we find a set
of operators that matches the DLCQ string spectrum in the pp-wave
background, which carry {\it discrete} light-cone momenta label by an
integer $k$ as well as discrete winding numbers $m$.  

To explain the basic construction and to simplify the notation we will
first discuss the states in the $k=1$ DLCQ sector. These states
correspond to a single ``bit'' of string carrying the smallest
possible momentum, namely $2p_+={1\over R_-}$.  After that we explain
how to put these string bits together to form strings with arbitrary
light-cone momentum.

\newsubsection{DLCQ ground states and zero mode oscillators}

The simplest state in the string theory is the $|k=1,m=0\rangle$ DLCQ
ground state.  According to our identification of $p^+$ this should be
a state with $\Delta=N_2J+J'=N_2$.  Since this state has $H=0$, it is
clear that it has to be constructed out of the $A_I$ alone.  The $A_I$
fields are bi-fundamental with respect to the pair of gauge groups
$SU(N_1)^{(I)}\times SU(N_1)^{(I+1)}$, hence the simplest
gauge-invariant operator that can be made out of them should contain
all $N_2$ $A_I$ fields precisely once. We thus arrive at the
identification
\be
|k\!=\!1,m\!=\!0\rangle = {1\over \sqrt{\cal{N}}}
\trsm(A_1 A_2\cdots A_{N_2}) 
\label{ground}
\ee
with ${\cal N}= N_1^{N_2}$. 
\figuren{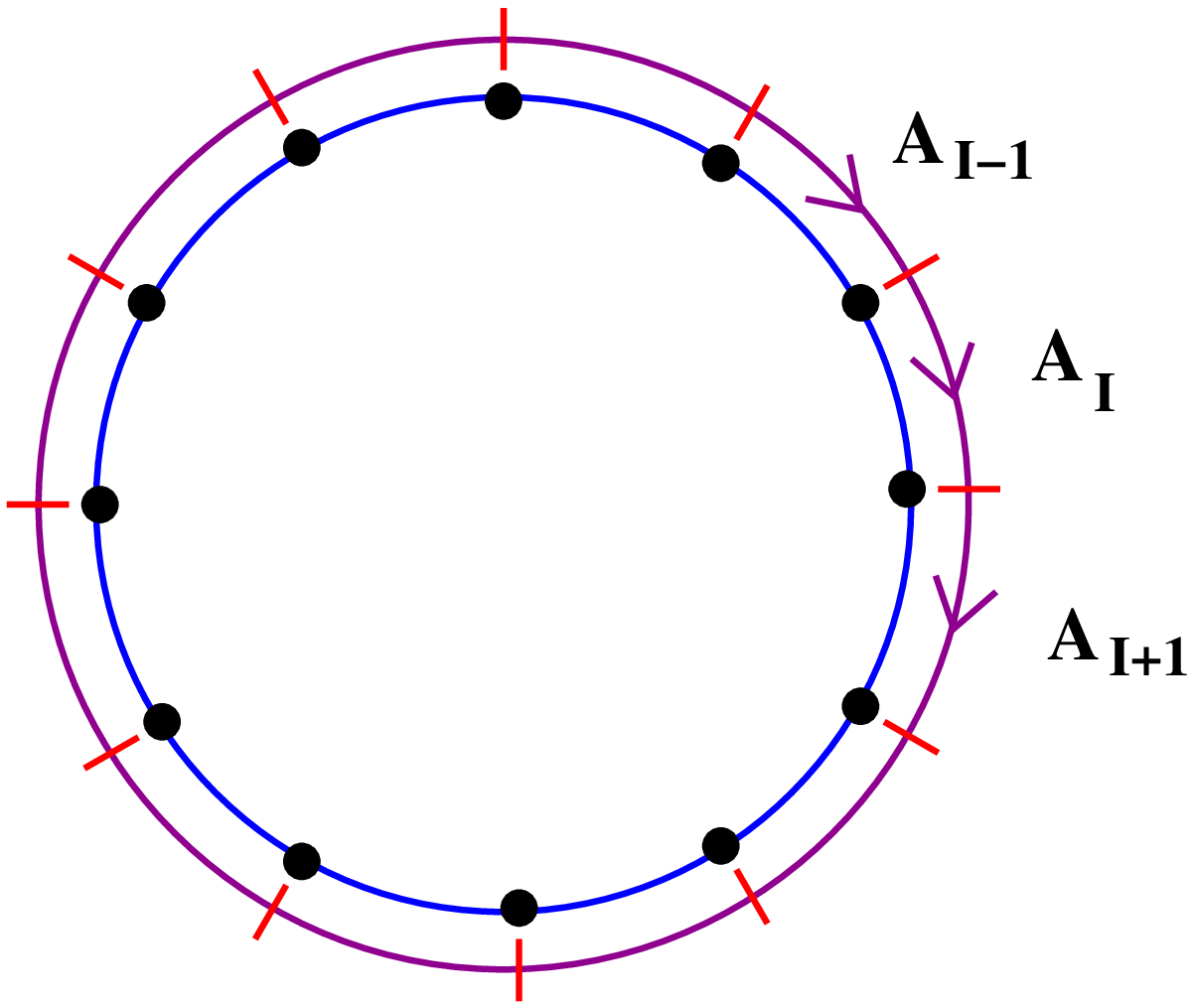}{7cm}{The $k=1$ ground state, $|k\!=\!1,m\!=\!0\rangle
\sim \trsm(A_1 A_2\cdots A_{N_2})$.}
This operator has $H=0$ and $\Delta=N_2$,
and is illustrated in Fig.\thefigure, where it appears as a string of
$A_I's$ wrapping once around the moose diagram.
The normalization of the operator can be determined by looking at the
two-point function in the free field limit. Basically, each of the
$A_I$ has to contract with the corresponding object and we get a
factor of $N_1$ from each gauge group trace. 

The states with arbitrary light-cone momentum are obtained by
literally stringing together the $k=1$ string bits. In particular, the
ground states in the sector with $k$ units of momentum is described by
\be
{|k,m\!=\!0\rangle=
{1\over\sqrt{{\cal N}^k}}
\trsm \left({{\underbrace{A_1 A_2\cdots A_{N_2}A_1A_2\cdots 
A_{N_2}\ldots\ldots\ldots A_1A_2\cdots A_{N_2}}}}\right). 
\atop{\mbox{ $\qquad\qquad\qquad\quad k$ times}}}
\label{kops}
\ee
Thus we have a single gauge-invariant operator with $H=0$ and
$\Delta=N_2 k$ for each value of $k$. One easily checks that it has the
right properties to describe the DLCQ ground state with momentum
$2p^+={k\over R_{-}}$. 
\figuren{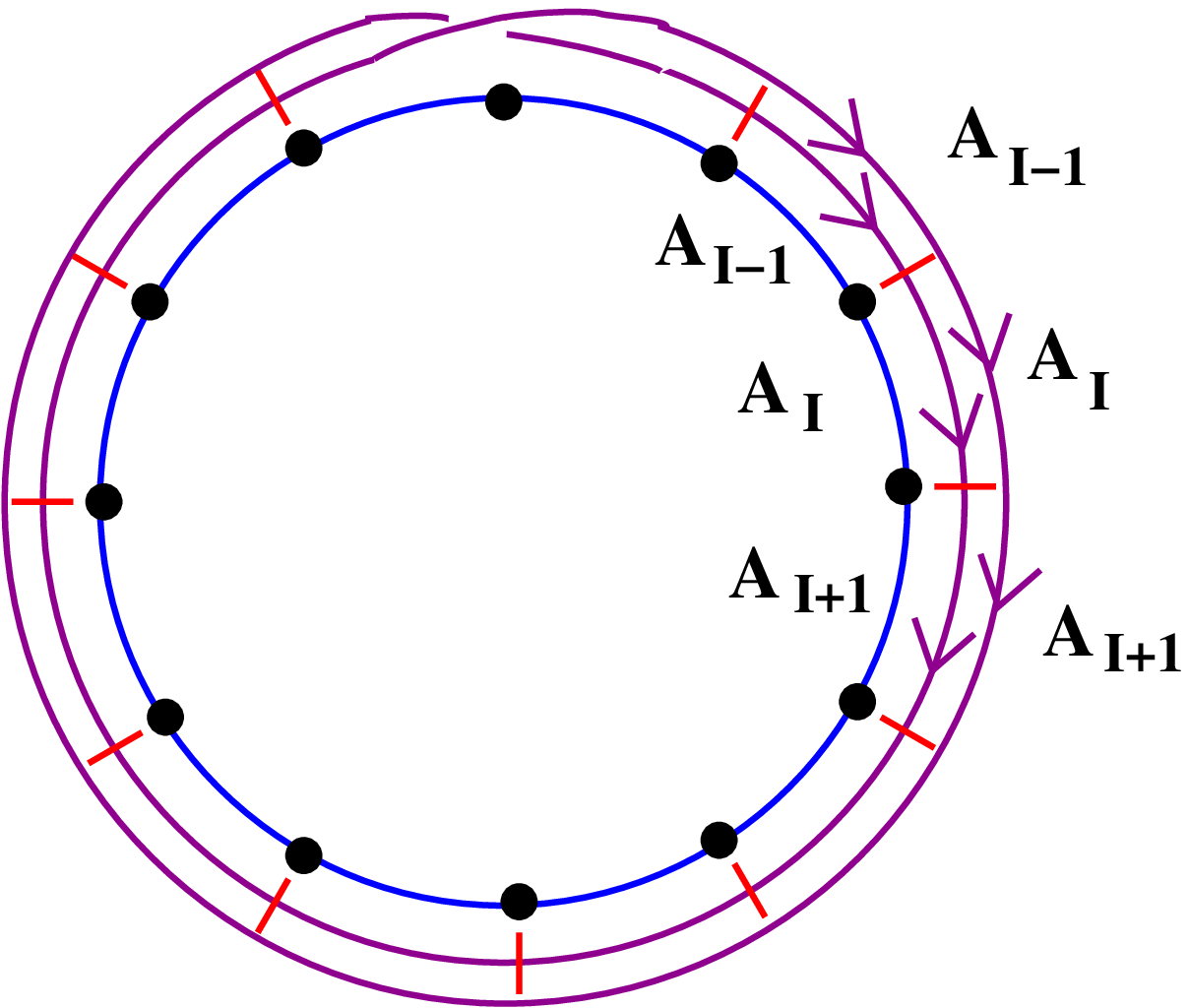}{7cm}{The $k=2$ ground state, $|k\!=\!2,m\!=\!0\rangle$
$\sim$ $\trsm(A_1 A_2\cdots A_{N_2}A_1 A_2\cdots A_{N_2})$.}
The $k=2$ operator is pictorially represented in Fig.\thefigure\ as a
string of $A_I$'s that wraps twice around the moose.

String oscillator modes are obtained by
inserting the various fields with $H=1$ in the appropriate locations
in the string of $A$-fields. We will first discuss the oscillator
states for the case $k=1$. In this case the notation is simpler, and
furthermore, it will more clearly exhibit the differences between the
operators in the quiver theory and the BMN-operators of the 
${\cal N}=4$ Yang-Mills theory.

We now come to the first excited states in the $k=1$ sector with
light-cone hamiltonian $H=1$.  For the string in the pp-wave
background, these states are obtained by acting once with a single
zero mode oscillator on the ground states. There are eight bosonic
zero mode oscillators, corresponding to the transversal coordinates
$x^i$, so we expect to find eight bosonic states with $H=1$.  In the
gauge theory these states are obtained by inserting appropriate
combinations of the $\Phi$ and $B$ fields and covariant derivatives
into the trace of the string of $A$-fields.

{}From the dimensions and charges of the bosonic operators listed in
the table, it is clear that we can admit precisely one insertion of a
covariant derivative, a $\Phi_I$, $B_I$, or their complex conjugates,
since they all have $H=1$. The matrix nature of these fields (adjoint
or bi-fundamental) constrains what gauge-invariant operators can be
written down.  For example, the fields $\Phi_I$ are in the adjoint of
$SU(N_1)^{(I)}$ and therefore must be inserted between $A_{I-1}$ and
$A_I$ in the string of operators.  The same is true for ${\bar\Phi}_I$
and the covariant derivatives $D_i^{(I)}$.  The field $B_I$, however,
is a bifundamental and requires an extra insertion of $A_I$, while
${\overline{B}}_I$ can only be inserted in the place of $A_I$.

In this way we get the following set of operators with $H=1$. First,
for the $\Phi$ fields we have
\bea
a^\dagger_{\Phi,0}|k\!=\!1,m\!=\!0\rangle & = 
& {1\over \sqrt{N_1N_2{\cal N}}} \sum_{I=1}^{N_2}\trsm 
(A_1 A_2\cdots A_{I-1} \Phi_I A_I \cdots
 A_{N_2}) \\
a^\dagger_{\overline{\Phi},0}|k\!=\!1,m\!=\!0\rangle 
& = & {1\over \sqrt{N_1 N_2{\cal N}}}
\sum_{I=1}^{N_2}\trsm (A_1 A_2\cdots A_{I-1} 
{\bar\Phi}_I A_I \cdots A_{N_2})
\eea 
The expressions for the covariant derivatives $D_i^{(I)}$ are similar,
and will not be written explicitly. The states containing $B$ fields are:
\bea
a^\dagger_{B,0}|k\!=\!1,m\!=\!0\rangle 
& = & {1\over \sqrt{N_1^2 N_2{\cal N}}} \sum_{I=1}^{N_2}\trsm 
(A_1 A_2\cdots A_{I} B_I A_I \cdots
 A_{N_2}). \\
a^\dagger_{\overline{B},0}|k\!=\!1,m\!=\!0\rangle 
&=& {1\over \sqrt{N_2{\cal N}}}
\sum_{I=1}^{N_2}\trsm (A_1 A_2\cdots A_{I-1} 
{\overline{B}}_I A_{I+1} \cdots A_{N_2})
\eea
The sum over the position of the insertions is necessary to ensure
that we are describing the zero mode fluctuations. This differs from
the BMN operators $\trsm(Z^J\Phi)$ for which the sum was automatically
implemented by the cyclicity of the trace. Thus we have found the
eight expected bosonic states at $H=1$. 
\figuren{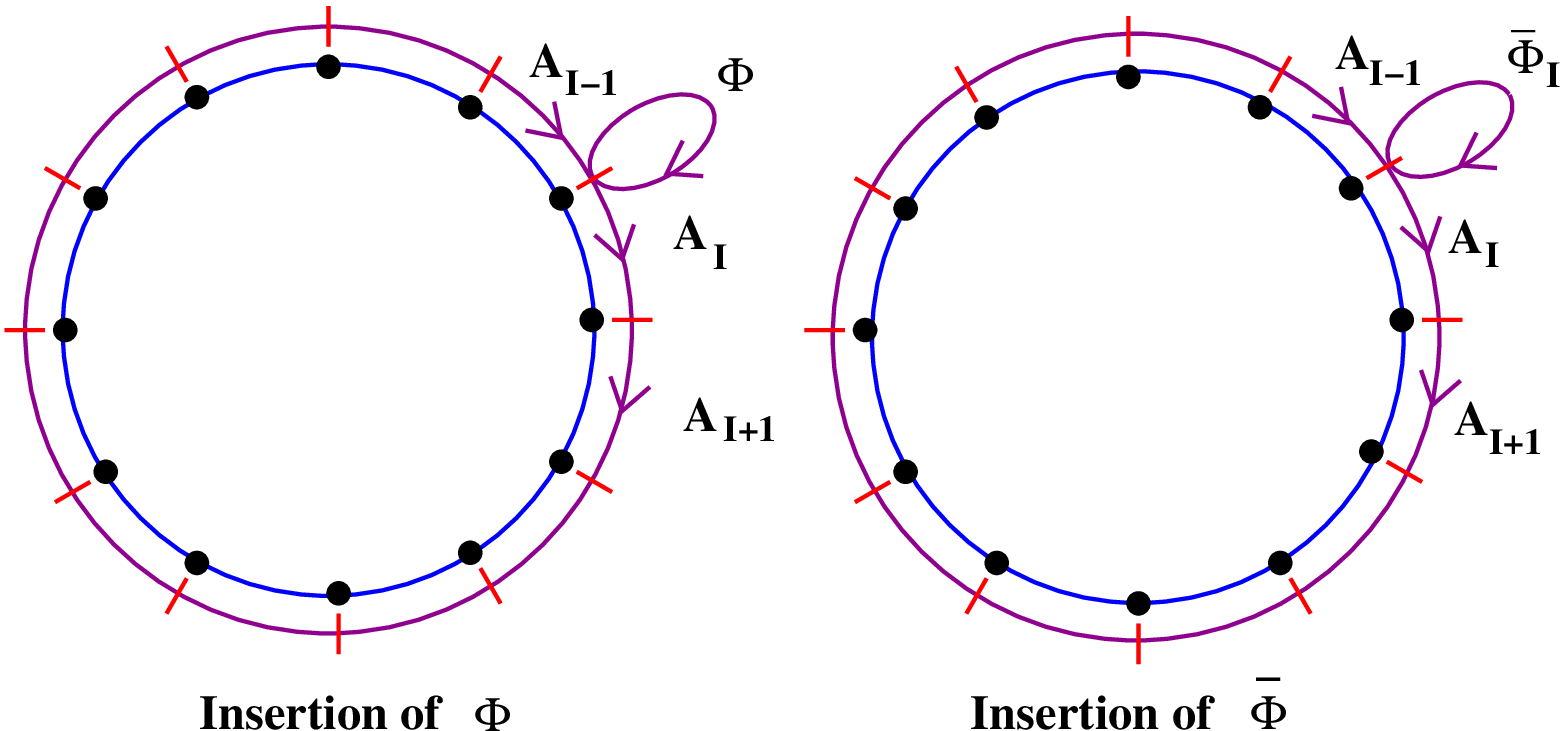}{12cm}{The building blocks for the states
$a^\dagger_{\Phi,0}|k\!=\!1,m\!=\!0\rangle$ and
$a^\dagger_{\overline{\Phi},0}|k\!=\!1,m\!=\!0\rangle$.} 
The operators involving insertions of $\Phi,\bar\Phi$  
are represented in Fig.\thefigure,
\figuren{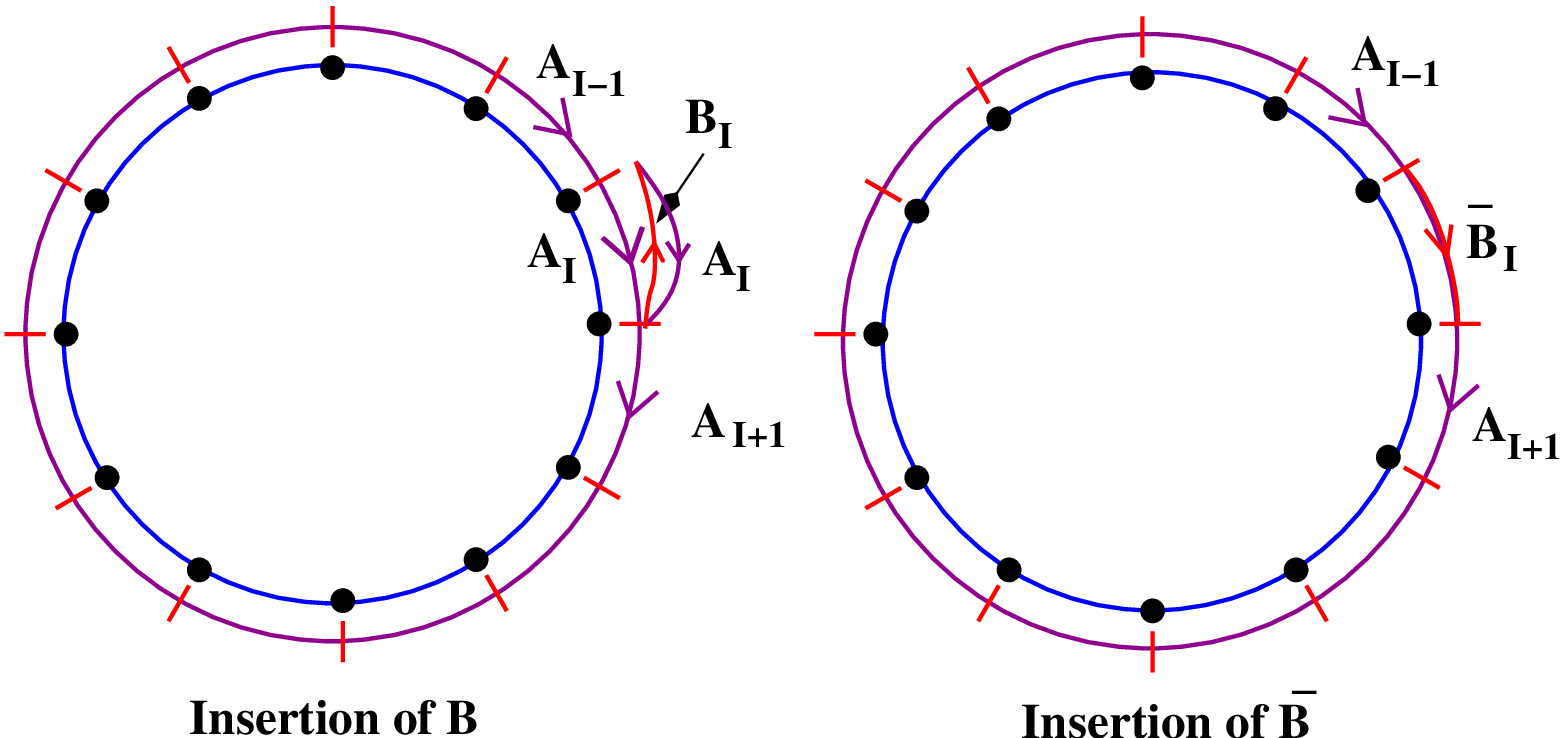}{12cm}{The building blocks for the states
$a^\dagger_{B,0}|k\!=\!1,m\!=\!0\rangle$, 
$a^\dagger_{\overline{B},0}|k\!=\!1,m\!=\!0\rangle$.}%
and those representing insertions of $B,\bar B$ in Fig.\thefigure.

In addition there are eight fermion states.  They are obtained in a
similar way by inserting the fermionic partners of these fields.  The
complex $A$, $B$ and $\Phi$ scalars each have a Weyl doublet fermionic
partner, that together make up 12 real fermionic fields. In addition
we have two gaugino fields, making up 16 fermions in the
theory. {}From the table, we see that precisely half of these have
$H=1$, and these are the ones that are used to construct the eight
fermionic oscillators of the string. Four of these are associated to
$\chi_A$ and ${\bar\chi}_B$, which are superpartners of the $A$ fields
under ${\cal N}=2$ supersymmetry.

\newsubsection{Winding states and non-zero mode oscillators}

An important difference between our large moose/quiver theory and the
${\cal N}=4$ Yang Mills theory is that the $A$-fields carry labels,
and have to appear in a particular order inside the trace.  This means
that a field that is inserted somewhere in the string of $A$'s will
have a definite position associated with it which can not be changed
using cyclicity of the trace.  This fact will be important in what follows 
because it allows us to identify the states with non-zero
winding along the light-cone.  To keep the notation simple we consider
insertions of only $\Phi$ fields. The extension of our discussion to
other fields will be straightforward. We also drop the normalization
factors from here on.

Similarly as in Ref.\cite{BMN} we can construct operators in which the
sum over the locations of the fields includes phase factors. The
simplest states of this kind are the single oscillator states with
winding number $m$
\be
a^\dagger_{\Phi,m}|k\!=\!1,m\rangle = 
\sum_{I=1}^{N_2}\trsm 
(A_1 A_2\cdots A_{I-1} \Phi_I A_I \cdots
 A_{N_2}) \omega^{mI},\\
\ee 
where $\omega = e^{2\pi i\over N_2}$.  Note that, unlike the cases
previously studied, these operators do not vanish due to the cyclicity
of the trace. This is just as well, because we want to have these
states in the DLCQ spectrum of the string.  More generally we can
construct operators with multiple insertions of $\Phi$-fields
corresponding to states with more oscillators.
\be
{}\!\!\prod_{i=1}^M a_{\Phi,n_i}^\dagger|k\!=\!1,m \rangle
= \sum_{l_M>\cdots>l_2>l_1}^{N_2}\!\!\trsm(A_1\cdots A_{l_1-1} 
\Phi_{l_1}A_{l_1}\cdots
A_{l_i-1} \Phi_{l_i}A_{l_i}
\cdots A_{N_2})\,\omega^{\sum {n_i l_i}}
\ ,
\label{manyins}
\ee
where the winding number $m$ is defined as the sum of the mode numbers
$n_i$:
\be
 m\equiv\sum_i n_i.
\ee
The operators (\ref{manyins}) represent the most general perturbative
string states in the $k=1$ sector with one unit of light-cone
momentum.

\newsubsection{Matrix notation}

Before we turn to the operators with general $k$, we would like to
rewrite our result for $k=1$ in a concise notation. For this
purpose, it will be convenient to introduce $(N_1N_2)\times (N_1 N_2)$
matrices to represent the $(\Phi_I, A_I, B_I)$ as follows:
\be
\mathbf A \equiv \pmatrix{0 &A_1& 0&\cdots &0\cr 0& 0 &A_2 &\cdots
&0\cr
\vdots& &  & \ddots&\vdots\cr
0 & 0 & 0 &\cdots &A_{N_2-1}\cr
A_{N_2}& 0 & 0 &\cdots &0\cr} \qquad
\mathbf B \equiv \pmatrix{0 &0 & \cdots& 0 &B_{N_2}\cr
B_1& 0 &\cdots &0 &0\cr
0 & B_2 & \cdots &0 &0\cr
\vdots& & \ddots &\vdots &\vdots\cr
0& 0 & \cdots &B_{N_2-1} &0\cr} 
\ee
and
\be
{\bphi} \equiv \pmatrix{\Phi_1 &0 &\cdots &0\cr
0 &\Phi_2  &\cdots &0\cr
\vdots & &\ddots & \vdots\cr
0 &0 &\cdots &\Phi_{N_2}\cr}
\ , 
\ee
where the blocks $A_I,B_I,\Phi_I$ are $N_1\times N_1$ matrices. 

The $k=1$ ground state is simply $\tr(\mA^{N_2})$, where $\tr$
represents the trace over the larger $N_1N_2$-dimensional space. 
The eight bosonic states with $H=1$ may up to normalization be written
as
\be
\tr(\bphi\mA^{N_2}),  \tr(\bar\bphi\mA^{N_2}),  
\tr(\mB\mA^{N_2+1}),  \tr(\bar\mB\mA^{N_2-1}),
~ \!\tr (D_i\mA^{N_2}).
\ee 
In this notation the expression of our operators are very similar to
the BMN-operators $\trsm(Z^J\Phi)$ etc.  This is not a coincidence,
because our operators can be obtained from theirs via an orbifold
projection.  

In the sector with zero winding number we can write the states with
many oscillators as:
\be
\prod_{i=1}^M a_{\Phi,n_i}^\dagger |k\!=\!1,m\!=\!0\rangle
= \sum_{l_M>\cdots>l_2>l_1}^{N_2}
\tr(\mA^{l_1} \bphi \mA^{l_2-l_1} \bphi
\cdots \bphi \mA^{N_2-l_M})\,\omega^{\sum {n_i l_i}}
\ .
\label{manyinstwo}
\ee
{}It follows from cyclicity of the trace that the r.h.s.vanishes
unless $\sum_i n_i=0$ (mod $N_2$).  Cyclicity is implemented by the
shift $l_i\rightarrow l_i+1$, which causes the expression to pick up a
phase $\omega^{\sum {n_i}}$. This phase must be equal to 1 and hence
the total mode number is set to zero modulo $N_2$.  This all seems
nice and well, but we appear to have lost our winding states!  This is
not surprising because we have written our states in a notation that
is inherited from the parent ${\cal N}=4$ theory. The winding states
are not present in the parent theory, but originate as twisted sectors
in the orbifold.

To describe the states with non-zero winding we introduce the clock
matrix:
\be
\mV \equiv \omega \pmatrix{1 &0 &\cdots &0\cr
0 &\omega  &\cdots &0\cr
\vdots & &\ddots & \vdots\cr
0 &0 &\cdots &\omega^{N_2-1}\cr}
\ .
\ee
where the additional phase is introduced for convenience.
This is actually an $N_1N_2\times N_1N_2$ matrix made up of $N_1\times
N_1$ blocks proportional to the identity. It obeys
\be
\mA \mV=\omega\,\mV \mA
\ 
\label{avrel}
\ee
and commutes with $\Phi$.  To obtain the operators in the sector of
winding number $m$, we insert an explicit factor of $\mV^m$ in the
trace, say, at the end.  As a result a cyclic permutation of all the
other operators in the trace produces an extra phase $\omega^m$.
Hence the argument that first gave a zero total mode number now gives
indeed that $\sum_i n_i=m$.  In fact, we can use the clock matrix $V$
to rewrite the states with many oscillators in a way that does not
require any explicit phases to be inserted.  Namely, we associate to
each oscillator a matrix-valued operator given by:
\be
a_{\Phi,n}^\dagger \leftrightarrow \bphi_n \equiv \bphi \mV^n 
\ .
\ee
The matrix on the r.h.s. is diagonal and has entries
$\Phi_I\omega^{nI}$.  Then the general oscillator state for momentum
$k=1$ is simply given by:
\be
\prod_{i=1}^M a_{\Phi,n_i}^\dagger|k\!=\!1,m\!=\!\sum_i n_i\rangle
=\sum_{l_M>\cdots>l_2>l_1}^{N_2}\tr(\mA^{l_1} \bphi_{n_1} \mA^{l_2 -l_1}
\bphi_{n_2} \cdots \bphi_{n_M} \mA^{N_2-l_M})
\label{oscstate}
\ ,
\ee
where no additional phases are inserted. The winding number of this
state is the total ``mode number'' of the $\bphi_n$, which is just the
number of $\mV$ matrices inside the trace.

For future purpose we also introduce the shift matrix
\be
\mU\equiv \pmatrix{0 & 1& 0&\cdots &0\cr
0& 0 &1 &\cdots &0\cr
\vdots& &  & \ddots&\vdots\cr
0 & 0 & 0 &\cdots &1\cr
1 & 0 & 0 &\cdots &0\cr}
\ .\ee
The clock and shift matrices obey the familiar relation
$$
\mU\mV=\omega \mV\mU.
$$ 
We can for example use the shift matrix $\mU$ to shift all $A_I$
fields in the $\mA$ matrix by the map $\mA\to\mU^{-1}
\mA\mU$. Applying this map to the $\mA$ and $\bphi$ 
matrices in the operators
(\ref{oscstate}) gives back the same operator, but multiplied by a
phase $\omega^m$ where $m$ is the winding number.  This observation
will be useful when we consider the generalization to arbitrary 
light-cone momentum $k$.

\newsubsection{String states at arbitrary light-cone momentum}

To describe the operators for general $k$ we make use of a similar
matrix notation as just described for the $k=1$ states.  To this end
we introduce even bigger $\mA,\mB$ and $\bphi$ matrices of size
$kN_1N_2$, where for example
\be
\mA \equiv \pmatrix{0 &A_1& 0&\cdots &\cdots &\cdots &\cdots &\cdots &0\cr
0& 0 &A_2 &\cdots &\cdots &\cdots &\cdots &\cdots &0\cr
\vdots& &  & \ddots& & &\vdots &\vdots & \vdots\cr
0 & \cdots & \cdots &0 &A_{N_2-1} &0 &\cdots & \cdots &0\cr
0 & \cdots & \cdots &\cdots &0 & A_{N_2} &0 &\cdots &0\cr
0 & \cdots & \cdots &\cdots &\cdots & 0 &A_1 &\cdots &0\cr
\vdots& &  & \vdots& & &\vdots &\ddots &\vdots\cr
0& \cdots & \cdots &\cdots &\cdots &\cdots &\cdots &\cdots &A_{N_2-1}\cr
A_{N_2}& 0 & \cdots &\cdots &\cdots &\cdots &\cdots &\cdots &0\cr} 
\ee
where the string of entries above the diagonal is repeated $k$ times
before terminating in the lower left corner. We continue to use the 
previous notation to label these larger matrices.

The ground state in the momentum $2p^+= {k \over R_-}$ sector can simply be
written as $\tr(\mA^{N_2k})$.  Then, in a similar way as for $k=1$, we
can introduce $N_2k\times N_2k$ shift and clock matrices $\mU$ and
$\mV$, where $\mU$ again has unit entries just off the diagonal and in
the left lower corner, while 
\be
\mV={\rm diag}\,\Big(\omega^{1\over k},
\omega^{2\over k},
\cdots, \omega^{(N_2 k-1)\over k}\Big)
\ .
\ee
These matrices now satisfy: 
\be
\mU\mV = \omega^{1\over k}\mV\mU
\ .
\ee

The general oscillator states can be written as in Eq.(\ref{oscstate}) in
terms of the new $\mA, \bphi$ and $\mV$. We get
\be
\prod_{i=1}^M a_{\Phi,n_i}^\dagger|k,m\rangle
=\sum_{l_M>\cdots>l_2>l_1}^{N_2k}\tr(\mA^{l_1} \bphi_{n_1} \mA^{l_2 -l_1}
\bphi_{n_2} \cdots \bphi_{n_M} \mA^{N_2k-l_M})
\label{oscstates}
\ ,
\ee
In this case, the sum of all oscillator mode numbers is equal to
\be
\sum_i n_i=km.
\ee
To see that the l.h.s. is a multiple of $k$, we note that the
$\mA$ and $\bphi$ matrices are invariant under shifting all entries
over $N_2$ places.  Thus, we have
\be
\mU^{-N_2} \mA\mU^{N_2}=\mA,\qquad \mU^{-N_2}\bphi\mU^{N_2}
= \bphi
\ .
\label{Nshift}
\ee
We now apply the first relation to replace all the $\mA$ matrices
inside the trace. By conjugating the $\mU$ matrices through the $\mV$
and $\bphi$ matrices one picks up a phase 
$\omega^{N_2\textstyle \sum_i {n_i\over k}}$. 
In effect we have not changed anything: we simply shifted all $A_I$
fields by $N_2$ steps, which gives back the same operator.  Therefore,
the operator vanishes unless the phase is equal to one. Thus, one
concludes that the sum of all mode numbers is indeed a multiple of
$k$.

In most of the above discussion, we have focussed on the oscillators
constructed out of $\bphi$.  However, it is straightforward to see
that similar expressions hold for the remaining oscillators, with
$\bphi$ replaced by $\bar\bphi$, $\mA\mB$, $\mA^{-1}\bar\mB$ and
$D_i$, or one of the fermionic fields.  This completes our
construction of perturbative string states for the pp-wave background
with a compact light-cone direction. The entire spectrum of the string 
in the DLCQ pp-wave is thus reproduced by this set of operators of the 
${\cal N} = 2 $ gauge theory. They represent a sector of the gauge theory 
with maximal supersymmetry.

The picture of gauge theory operators winding around a circular moose
diagram, which has a circumference of order $N_2$ and is therefore
very large, is rather suggestive. Even though these operators describe
momentum modes in type IIB string theory, they appear to build up a
string winding over a spatial dimension, Similarly, the gauge theory
operators describing winding modes in type IIB string theory have the
appearance of momentum states. This strongly suggests that T-duality
is involved. In the next section we will carry out the relevant
T-duality explicitly and exhibit the relation between the large moose
diagram and the T-dual string.

This is also somewhat related to the idea of ``deconstruction''
\cite{Arkani,Pokorski, ACKKM, Rothstein} in which a spatial dimension
is created by taking a suitable limit of a moose/quiver
theory. However, as we will explain in section 9., our limit differs
from the one in Ref.\cite{ACKKM}.

\newsection{T-duality of the DLCQ PP-wave Background}

The periodicity of the $x^-$ direction is a remnant of the combined
$2\pi\over N_2$ periodicity in the angles $\chi$ and $\phi$, exhibited
in Eq.(\ref{periodic}). Before taking the limit $N_1,N_2\to \infty$ the
periodic direction was space-like. Hence, one can perform a 
Buscher-type T-duality along this periodic direction.  

To perform the $T$-duality, we go back to the original metric in
Eq.(\ref{adsmet}) and write down only the terms in the $t$ and $\chi$
directions explicitly:
\be
ds^2 = R^2\left[ - \cosh^2\rho \, dt^2 + \cos^2\alpha \cos^2\gamma\,
d\chi^2\right] + ds_{\rm transverse}^2
\ .
\ee
Next we make the substitutions defined in Eq.(\ref{pplimit}),
and express $\chi$ in terms of $t$ and $x^-$ as in Eq.(\ref{lcone}).
As a result, the metric becomes:
\bea
ds^2 &=& R^2 \left(\cos^2{w\over R}\cos^2{y\over R}-\cosh^2 {r\over
R}\right) dt^2 - 4 \cos^2{w\over R}\cos^2{y\over R}\, dt\, dx^-
\nonumber\\
&&+ {4\over R^2} \cos^2{w\over R}\cos^2{y\over R}\, (dx^-)^2
+ h_{ij}\left({x^i\over R} \right) \, dx^i dx^j
\ ,
\eea
where we parametrize the transverse metric by $h_{ij}$. We know that
$h_{ij} \left({x^i\over R} \right) \to \delta_{ij}$ as $R\to\infty$.

We see that, before taking the pp-wave limit, $x^-$ is spacelike:
there is a small $g_{--}$ in the metric of order ${1\over R^2}$. The
metric of the transverse space, which we have not written down
explicitly here, is the usual flat metric with no factors of $R^2$ in
front. For the transverse space we can consistently ignore corrections
in ${1\over R^2}$, since those directions will be unaffected by
T-duality.

We now perform a T-duality along the spacelike direction $x^-$,
following the usual duality rules. Note that this is a different 
T-duality than the one considered in \cite{CLP}. After this, we denote the
T-dual coordinate by $2x^9$, to end  up with the metric:
\be
ds^2 = -R^2  \cosh^2 {r\over R}\, dt^2
+ {R^2\over \cos^2{w\over R}\cos^2{y\over R}} 
(dx^9)^2 + h_{ij}\, dx^i dx^j
\ ,
\label{tdualmetric}
\ee
along with a B-field and dilaton:
\be
B_{t\,9} = -R^2,\qquad g^A_s = {\sqrt{\alpha'}
R\over R_-\cos{w\over R}\cos{y\over R}}
\, g^B_s
\ .
\label{tdualb}
\ee
Note that $x^9$ now has period ${2\pi \alpha'\over R_-}$, with
$R_-$ as given in Eq.(\ref{rminus}).

Now let us take the limit $R\to\infty$. We see that some components
of the metric, and the $B$-field and string coupling, become infinite
in this limit. However, it turns out that string propagation on this
background is finite. The reason is that the $B$-field is a critical
electric field, and cancels the leading divergent piece in the string
world-sheet action $\sqrt{-\det(\gamma)} + B$ where $\gamma_{ab}
= g_{ab} + \partial_a x^i\partial_b x^i$ is the induced metric in
static gauge. Here $g_{ab}$ denotes the spacetime metric in the
$(t,x^9)$ plane. Momentarily ignoring the derivative terms, we find:
\bea
\sqrt{-\det(g)} + B &=& R^2{\cosh{r\over R}\over 
\cos{w\over R}\cos{y\over R}}  - R^2\nonumber \\
&=& \frac12 (r^2 + w^2 + y^2) + {\cal O}\left({1\over
R^2}\right)\nonumber \\
&=&  \frac12 \sum_{i=1}^8 (x^i)^2 + {\cal O}\left({1\over
R^2}\right)
\ .
\eea
If we put back the derivative terms, the above calculation will
instead give:
\be
\sqrt{-\det(\gamma)} + B =  \frac12 \sum_{i=1}^8\left[ 
\partial_a x^i \partial^a x^i + (x^i)^2 \right] +
{\cal O}\left({1\over R^2}\right).
\ee
This is just the non-relativistic string propagating in a background
with a Newtonian potential of harmonic-oscillator type. Indeed, the
leading dependence on $R$ in Eqs.(\ref{tdualmetric}, \ref{tdualb})
above is identical to that which appears in the Non-Commutative Open
String (NCOS) \cite{SST,GMMS} and Non-Relativistic Closed String
(NRCS) theories \cite{klebmal,gomisnr,danielsson}.  Our model is
therefore an NRCS theory. However, because of the pp-wave metric that
we started with, it inherits a harmonic oscillator potential in which
the light closed strings are bound.

To see that the effective coupling constant of this non-relativistic
theory is finite, we recall that the coupling for the NR
closed strings winding in the direction of the critical electric field
is the same as for the NC open strings. The latter is well-known to
be \cite{ACNY}:
\be
g_o^2 = g_s \sqrt{\det(g+B)\over {\det g}}.
\ee
Inserting the appropriate values from Eqs.(\ref{tdualmetric},
\ref{tdualb}) above and taking $R\to\infty$, we see that for our model,
\be
g_o^2 =  {g^B_s \over R_-}\sqrt{\alpha'\,\sum_{i=1}^8 (x^i)^2}.
\ee
So the effective coupling for the type IIA NR closed string is
independent of $R$ in the limit, but it varies in the transverse
directions. Because of the potential, low-lying states of the string
are localized in the region around $x^i\sim 0$, and on dimensional
grounds one expects that the square root in the above equation is
replaced by $\alpha'$ for such states.

There is also a 5-form Ramond-Ramond background on the type IIB side
that must be T-dualized. To start with, we had:
\be
F^{(5)} = R^4 \, dt\wedge (dV)_4 + R^4\, d\chi\wedge (dV')_4
\ee
where $dV,dV'$ are the 4-forms on the $(\rho,\Omega_3)$ and
$(\alpha,\theta, \gamma,\phi)$ spaces respectively. We use
Eq.(\ref{pplimit}) and the definition of $x^-$ in Eq.(\ref{lcone}),
and take the limit $R\to\infty$. Then the above expression becomes:
\be
F^{(5)} = dt\wedge (dr)_4 + dt\wedge (dw)_2\wedge (dy)_2
- {2\over R^2} dx^- \wedge  (dw)_2\wedge (dy)_2 
\ .
\ee
Next, performing a T-duality, we find a 4-form and a 6-form
field strength on the type IIA side:
\bea
F^{(6)} &=& 2 dt\wedge (dr)_4 \wedge dx^9,\nonumber \\
F^{(4)} &=& -{2\over R^2} (dw)_2\wedge (dy)_2
\label{rrforms}.
\eea
Thus we have an electric 6-form and a magnetic 4-form field
strength. As one can check, the two are dual to each other. These RR
fields will give masses to the worldsheet fermions of the string,
which are necessary for maximal supersymmetry, just
as the 5-forms do on the type IIB side. 

In section 8 we discuss the lift of this background to
M-theory, on which it gives rise to an analogous non-relativistic
membrane wrapped on a 2-torus.

\newsection{Large Quiver as Non-relativistic String}

We have argued that the quiver gauge theory with gauge group
$SU(N_1)^{N_2}$ with large $N_1,N_2$ is dual to type IIB string theory
on the pp-wave background with a compact lightlike direction $x^-$. We
have also shown that one can perform a T-duality over the lightlike
direction, thereby going to a type IIA description in terms of a
non-relativistic closed string theory (NRCS). In the following, we
present a number of observations which address the physical meaning of
this correspondence. We will argue that the large moose/quiver theory
deconstructs the non-relativistic string, and eventually M-theory, in
a rather precise way. 

We have described how ground states and oscillators of type IIB string
theory on a DLCQ pp-wave background can be constructed from gauge
theory operators. A key ingredient was the construction of a string
ground state for every (positive integer) DLCQ momentum $k$.  This
state was associated to the trace of a ``string'' of bi-fundamental
$A$ fields wrapping $k$ times around the ``theory space'' defined by
the quiver diagram. Although in the type IIB language it is a momentum
state, in the quiver theory it is reminiscent of a winding state of
some other string theory. Similarly, the states describing the winding
of IIB strings around the DLCQ direction, obtained using the
$\mV$-matrices, look very much like momentum states of a string theory.

We now see that this other string theory is precisely the type IIA
NRCS theory! It is known that in NRCS theory, along the direction of
the critical electric field there are light states which are
closed-string winding modes. However, the winding can take place only
in one orientation around the circle. The physical reason is
simple \cite{klebmal}. Consider the Non-Commutative Open String (NCOS)
theories obtained by turning on critical electric $B$-fields over a
D-brane. In these theories, we know that open strings are very light
when they align one way along the electric field, and very heavy when
they are lined up the other way. Now if the direction along the
electric field is compact, an open string can align along it and
stretch most of the way around. When its end points come close to each
other, and because the open-string coupling is finite, this string can
close up and become a closed string wound on the compact direction. As
a result, NCOS theories on compact directions include light winding
strings. If now we remove the open-string sector by taking away the
original D-brane, we will be left with the closed strings wound in one
direction as the light states. This is the NRCS theory.

Evidently these NRCS winding modes are just the T-dual states of the
DLCQ momentum modes. Just as DLCQ momenta are always positive and
never negative (the latter become infinitely heavy), the NRCS winding
modes are light for one orientation and very heavy for the other.
Thus winding states of the non-relativistic closed string are identified
with the gauge theory operators wrapping our large moose/quiver. 

In fact, the quiver deconstructs this string in a very precise
way. The ``string'' of $A$-fields winding around the quiver in one way
is the light NR closed string. A string of $\bar A$-fields, which wrap
the quiver in the other sense, would be identified with the very heavy
closed string that is wrongly aligned with the electric field, so it
is not in the spectrum of light states. Single wrapping of $A_I$,
corresponding to $k=1$, is identified with a ``string bit'' in DLCQ
language, while the $A_I$'s themselves are the latticized components
of this string bit. The vanishing $p^-$ value of the $A$ fields leads
to the vanishing energy of the string bit which is aligned with the
electric field. And the $\mV$ operators in the quiver theory, which
created winding states of the DLCQ string, are the momenta of the NR
closed string. Indeed, we can insert $\mV$ raised to any positive or
negative integer power, just like the allowed momenta of an NRCS.  The
quiver theory in the limit we consider is a non-relativistic string
with a quadratic potential. The observations of \cite{klebmal,
gomisnr,danielsson} made in the context of flat space NRCS theory and
its relation to DLCQ strings are not modified in the pp-wave
background.

\newsection{Beyond Perturbation Theory}

So far we have analyzed the large moose theory at the perturbative
level. In this section we will discuss some ideas on how to describe
non-perturbative aspects of our theory. In particular, we propose an
identification of D-string states in terms of operators in the gauge
theory. We also discuss the lift of the type IIA solution to
M-theory. 

\newsubsection{D-string States}

In addition to the fundamental string oscillator and winding states
that we have discussed, we also expect to find states in the string
theory that represent D-strings winding around the DLCQ
direction. These would be S-dual, in the framework of type IIB string
theory, to the F-string winding states. In fact, using various
elements of the $SL(2,\mZ)$  S-duality group, one can construct $(p,q)$
winding strings where $p$ and $q$ are relatively prime.

The moose/quiver gauge theory that we started with is also believed to
have a large S-duality group which contains $SL(2,\mZ)$ as a subgroup 
\cite{Wittenm}. The perturbative operators
that we have considered so far are constructed out of electric variables
that are weakly coupled when the Yang-Mills coupling is small. But in
principle for each operator that we have constructed there must exist
corresponding operators that are expressed in magnetic or even in 
dyonic variables. These different operators are then related by 
$SL(2,\mZ)$ tranformations. The ground states that we considered can 
easily be seen to be invariant under the electromagnetic $SL(2,\mZ)$  
transformations. They represent supersymmetric graviton states that 
carry only space-time momentum. Also the states that are
obtained by acting with zero mode oscillators will be invariant: they 
correspond to gravitons whose transversal movements are described by 
excited harmonic oscillator states.

The non-zero mode oscillator states and the states with non-zero winding 
are true $F$-string states, and will transform non-trivially 
under $SL(2,\mZ)$ and are mapped on excited $(p,q)$ string states with 
non-zero winding. It is useful to think about these states from a $T$-dual
perspective. After $T$-duality the discrete light-cone momentum $k$ 
becomes string winding, while string winding $m$ is mapped on to the
discrete momentum along the string. $D$-string winding becomes identified 
with $D$-particle number. In this dual language the extra states that 
we are looking for are thus described by strings bound to $D$-particles.

Can we introduce $D$-particles on the quiver/moose diagram?
 This is a hard question, because it involves 
non-perturbative issues in the gauge theory, and might require the introduction
of the dual magnetic gauge field. However, there appears to be 
a natural proposal for these $D$-string/particle states which only 
involves pure electric variables. To give some motivation for the proposal, 
we note that the perturbative string states represent ``electric'' states
associated to the confining phase of the theory. This suggests that
the ``magnetic'' states must be obtained from a dual Higgs
phase. Suppose the $A$-fields get a vacuum expectation
value. $\mZ_{N_2}$ invariance implies that they all get the same
VEV. Since the theory is conformal invariant, we can put these VEV's
equal to one. Hence we have $\langle \mA\rangle = \mU$.
In the Higgs phase, it is very natural to construct operators using
the shift matrix $\mU$ in the same way as we did for the $\mV$-matrix.
Specifically, in the $k=1$ DLCQ sector we can define oscillators
\be
a^{\dagger}_{_\Phi,m,n}\leftrightarrow \bphi_{m,n}\equiv 
\bphi \mU^m \mV^n 
\ ,
\ee
insert them into the trace, and sum over all locations. The F- and D-
string winding numbers may then be defined as the sum of the $n$ and $m$
mode numbers respectively.  The states with $n=0$ will be pure
D-strings, while those with $(n,m)=l(p,q)$ with $(p,q)$ both nonzero
and relatively prime will correspond to $l$ $(p,q)$ strings.  Clearly
this reproduces the operators we had before, and extends them in an
almost manifest $SL(2,\mZ)$  invariant manner. There is one subtlety
however. The number of $\mA$ fields that are inside the trace must be
reduced by the number of $D$-string windings, because otherwise the
trace would vanish. The $U$ matrices play the role of `place holders'
and create `holes' in the string of $A$-fields. This is just because
the fields are replaced by their vacuum expectation values.
These holes are the locations of the $D$-particles on the string.

This can be generalized to general $k$ values by extending 
$\mU,\mV,\mA,\bphi$ to be $(kN_1N_2)\times(kN_1N_2)$ matrices, and 
requiring that the resulting states be invariant.  The
$D$-string/particle  number is then defined in terms of the phase that is
picked up when we replace all $\mA$ fields by $\mV\mA\mV^{-1}$ and
similarly for the $\bphi$'s. This phase will be an $N_2$'th root of
unity, and from $\mU\mV=\mV\mU\omega^{1\over k}$ we get 
\be
\omega^{\sum {m_i\over k}}\equiv \omega^{{\tilde m}},
\ee 
where $\omega =e^{2\pi i\over N_2}$ and ${\tilde m}$ is the
D-string winding number.

We should perhaps emphasize that the proposed description of 
the $D$-string/particle states came from a heuristic argument and 
was not derived in a precise way. However, we are pretty sure that
there must be states that carry these quantum number, basically because of
the $SL(2, \mZ)$ symmetry of the underlying gauge theory. 
At finite coupling the gauge theory contains perturbative as well as
non-perturbative states, and therefore it should also give a non-perturbative
of the IIB string in the DLCQ background description, or the type IIA 
string in the dual background. If this is indeed true than we should conclude 
that the large quiver/moose theory also has a dual M-theory description 
in which the $F$- and $D$-string states combine to form the states of a 
membrane that wraps the $x^9$ as well as the $x^{10}$ M-theory circle.
In the next section we will describe the $M$-theory background that is 
dual to the DLCQ pp-wave and in which the membrane is living.

\newsubsection{Lift to M-Theory}

The type IIA background T-dual to the DLCQ pp-wave, described in
Eqs.(\ref{tdualmetric}, \ref{tdualb}) can be lifted to M-theory in the
standard way at finite $R$. It will be convenient to rewrite these
formulae as follows. Define:
\be
H(w,y) = {1\over g_s^B}{R_-\over \sqrt{\alpha'} R} \cos{w\over
R}\cos{y\over R}
\ .
\ee
In addition, we rescale $x^9$ so that it has periodicity
$2\pi\sqrt{\alpha'}$. The type IIA background is now
given by:
\bea
ds^2 &=& - R^2 \cosh^2{r\over R}\, dt^2 + {1\over (g^B_s)^2 H^2}
(dx^9)^2 + h_{ij}\,dx^i dx^j,\nonumber \\
B_{t\,9} &=& - {R^2\over R_-},\qquad e^\phi = {1\over H} 
\ .
\eea
This solution lifts to the following M-theory background:
\bea
{1\over (l_p)^2}
ds_{11}^2 &=&  H^{2\over 3}\left[-R^2 \cosh^2{r\over R}\, dt^2 
+ H^{-2} \Big({1\over (g^B_s)^2}(dx^9)^2 + (dx^{10})^2 \Big)
+ h_{ij}\, dx^i dx^j \right], \nonumber \\
C_{t\,9\,10} &=& - {R^2\over R_-}.
\eea
where $x^9$ and $x^{10}$ are periodic modulo $2\pi$, and $R^2$ in
this equation is just $\sqrt{4\pi g^B_s N_1 N_2}$ with no dimensional
factors. {}From the metric, we see that the ratio of the physical radii of 
$x^{10}$ and $x^9$ is $g^B_s$, as expected. 

Again we see that the background itself is singular as $R\to\infty$,
but membrane propagation is finite and
non-relativistic. Non-relativistic membranes are related to OM (open
membrane) theory \cite{GMSS} in a similar way as non-relativistic
strings are related to NCOS theories \cite{gomisnr,alberto}.  We can
compute the membrane action in static gauge, again ignoring derivative
terms to start with:
\bea
\sqrt{-\det(g)} + C &=& {1\over g^B_s H} R\cosh{r\over R} - {R^2\over
R_-}\nonumber \\
 &=& {1\over 2R_-}\sum_{i=1}^8 (x^i)^2 + {\cal O}\left({1\over R^2}\right)
\ .
\eea 
As for the string, the kinetic terms can be restored leading to a 
free action for the transverse scalars on the membrane world-volume. 
The background will also have a 7-form field strength
and its dual 4-form, following from Eqs.(\ref{rrforms}). These will 
give masses to the fermions of the membrane world-volume theory as
required by maximal supersymmetry.

\newsection{Relation to Deconstruction}

We have been working with the $SU(N_1)^{N_2}$ ${\cal {N}} =2$
supersymmetric quiver gauge theory for large $N_1,N_2$. The same
theory, for finite $N_1$ and large $N_2$, was the starting
point for deconstructing $(2,0)$ superconformal field theory in six
dimensions \cite{ACKKM}. 

The essential idea in deconstruction \cite{Arkani} is to start with a
four-dimensional gauge theory for which the ``theory space'' is large.
To be precise, one takes a large number of gauge groups and introduce
matter in non-trivial representations of two of the gauge groups
simultaneously; a prototypical example of this being the quiver gauge
theory under consideration.  Upon going to the Higgs branch of the
theory by giving VEVs to the matter in the bifundamentals, one can at
some intermediate energy scales recover approximate five-dimensional
gauge dynamics. The role of the fifth dimension is played by the
direction in the theory space.  In the deep infra-red the theory
recovers four-dimensional behaviour, since the Higgs VEV causes the
large gauge group to be broken down to the diagonal subgroup. The
five-dimensional gauge coupling and the lattice spacing are governed
by the Higgs VEV along with other parameters of the gauge theory.

In  \cite{ACKKM}, this idea was extended to supersymmetric gauge
theories in four-dimensions, and the quiver ${\cal {N}} =2$ theory was
the starting point for deconstruction of {\em two} extra
dimensions. One takes a set of $N_1$ D3-branes probing a
$\mC^2/\mZ_{N_2}$ orbifold. However, if $N_2$ happens to be large,
then the resulting wedge of $\mC^2$ is like a thin sliver. The
D3-branes are moved away from the orbifold point to a distance $d$. In
the dual gauge theory, this corresponds to giving a VEV to the
bifundamental hypermultiplets. The D3-branes are now in a
very thin cone, and being away from the tip, they essentially can be
considered to be in a cylindrical geometry where the radius of the
transverse circle is $\frac{d}{N_2}$. For large $N_2$, when this
radius is vanishingly small in string units, one can T-dualize and
work with D4-branes wrapping a dual circle of radius $R^A
=\frac{N_2 \alpha'}{d}$. The Type IIA coupling constant is $g^A_s
=\frac{N_2 \sqrt{\alpha'}}{d}$.

The limit considered in Ref.\cite{ACKKM} is to take $N_2$ large and
$\alpha'\rightarrow 0$ with $g_s^B$ and ${d \over N_2 \alpha'}$
fixed. In this limit the Type IIA coupling blows up, so we can lift
the configuration to M-theory. The M-theory circle has a radius
$R_{10}= \frac{g^B_s N_2 \alpha'}{d}$ and the 11-dimensional Planck
length is $l_P^3 = \frac{(\alpha')^2 N_2 g^B_s}{d}$. The Type IIA
four-branes have become M5-branes wrapping the $x^{10}$ circle.  We
are left with an M-theory background with $N_1$ M5-branes, which for
$l_p\to 0$ defines the $(2,0)$ theory with parameter $N_1$.

To summarize, the basic idea in Ref.\cite{ACKKM} is to keep $N_1$
and $g_s^B$ fixed, taking $N_2$ large while staying in the Higgs
branch. The states which survive in the gauge theory are of
energies ${\langle A \rangle \over N_2}$, where $\langle A \rangle$ is
the Higgs expectation value.

In our approach, we go over to the $AdS$ limit, $N_1 \rightarrow
\infty$, simultaneously with the ``deconstruction'' limit
$N_2\to\infty$.  In the $AdS$ background, we pick a circle along the
$\chi$ direction, which lies a distance $R$ away from the fixed
circle of the orbifold action. 
This is the analog of staying away a distance $d$ from the tip
of the cone in the discussion of \cite{ACKKM}.  But $d$ is not to be
identified with $R$ as such, since the energy scale ${R \over N_2
\alpha'}$ is vanishingly small in the limit we consider. 
In addition to staying away from the fixed circle, the pp-wave limit
also involves a boost along the $\chi$ direction. This introduces
another factor of $R$, and therefore we need to compare the scale ${d
\over N_2\alpha'}$ of \cite{ACKKM} with ${R^2 \over N_2\alpha'^{3/2}}$. 
We look at states which have energy of order the inverse of this
length scale.

Recall that the radius of our compact null direction is proportional
to $\sqrt{N_1\over N_2}$. We now see that the conventional $AdS$ limit
and the limit of Ref.\cite{ACKKM} probe opposite ends of the DLCQ
compactification moduli space, with the former corresponding to
$R_-\to\infty$ and the latter corresponding to $R_-\to 0$. By a double
scaling of $N_1$ and $N_2$, we have succeeded in retaining the DLCQ
radius as a tunable parameter.

\newsection{Discussion}

We end with some final comments and speculate about using the 
large moose/quiver theory as a definition of M-theory.

\newsubsection{Concluding Comments}

Our work indicates several interesting directions to explore. One such
direction is to consider ${\cal N}=1$ superconformal gauge theories
obtained by placing $N_1$ D3-branes at a $\mC^3/(\mZ_{N_2}\times
\mZ_{N_3})$ orbifold singularity. The holographically dual 
$AdS_5\times \S^5/(\mZ_{N_2}\times \mZ_{N_3})$ background naturally
admits three different pp-wave limits. Two of these are DLCQ pp-wave
backgrounds with an orbifold in the transverse space, hence they have
reduced supersymmetry. The third limit, which is more interesting,
gives a maximally supersymmetric pp-wave. The corresponding quiver
diagram is a two-dimensional discretized torus, and the gauge theory
operators wind along a diagonal of this torus, sampling all the
$N_2N_3$ points \cite{MRV}. 

An alternative description of the ${\cal N}=2$ quiver theory that we
studied in this paper is provided by a type IIA brane construction in
terms of NS5 and D4-branes \cite{Wittenm}. This construction lifts to
M-theory as an M5-brane wrapped on a Riemann surface. It would be
worthwhile to examine whether this approach, in our scaling limit,
can be related to the non-relativistic membrane theory that we have
exhibited here. 

Another interesting study, similar to the deconstruction idea, deals
with the matrix theory description of our quiver theory \cite{GS}. In
this work, a relation is found between $SU(N_1)^{N_2}$ and
$SU(N_2)^{N_1}$ quiver theories. This exchange of $N_1$ and $N_2$ is
reminiscent of the T-duality that we perform, and might have
implications for it.

Finally, we note that the DLCQ formalism is intended to facilitate the
study of string interactions in a controlled fashion. It is clearly
important to understand string interactions within the gauge
theory/pp-wave correspondence, and the framework provided in this
paper should be useful to address this issue.

\newsubsection{Might `M' Mean `Moose'?} 

In this paper we considered a particular limit of the ${\cal N}=2$ 
moose/quiver theory in which not only $N_1$ and $N_2$ are taken to be
larger, but at the same time one focusses on a particular sector of the
theory: only operators are considered for which 
$\Delta$ and $J'$ grow like $N_2$ and  $J$ and $\Delta-N_2J-J'$ are kept
finite. The latter conditions imply that
the operators are constructed mostly out of $A$ fields. In principle one
could have chosen to look at a different sector for which another
combination of $\Delta$ and the charges is kept finite leading to
operators that contain for example mostly $\bar{A}$, $B$ or $\bar{B}$
fields. This seems to suggest that the large moose/quiver theory may be
even richer and may contain all kinds of sectors that we have not yet
explored. 

This brings us to an important question: is one allowed to
send $N_1$ and $N_2$ to infinity without taking the string coupling to
zero? Or does this only makes sense provided one also restricts to an
appropriate subset of operators? To see why this is an important
question, let us assume that one {\it can} make sense of the large
moose/quiver theory  without any severe restrictions on the coupling or
the class of operators  one is considering. 
In that case one expects that the resulting theory is dual to
type IIB string theory on $AdS_5\times \S^5/\mZ_{N_2}$ in limit where the
radius of the $AdS_5$ and the $\S^5$ both go to infinity, while taking
also the order $N_2$ of the $\mZ_{N_2}$ orbifold group to infinity.
Without the orbifold identification one would, at least naively, expect
to get the type IIB theory in a flat background. One of the spatial
direction is, however, infinitely small due to the orbifold symmetry,
and hence should be 
replaced by a T-dual coordinate. This gives a type IIA theory in flat
space but with a coupling constant that becomes infinite. In other
words, the large $N_1$ and $N_2$ limit of the moose/quiver theory, if it
exists, is a serious candidate for a non-perturbative description of
type IIA string theory, that is of $M$-theory! So we may ask: ``Might M
mean Moose?''

\section*{Acknowledgements}

We thank Robbert Dijkgraaf, Veronika Hubeny, Igor Klebanov, Juan
Maldacena, Horatiu Nastase, Nati Seiberg, Savdeep Sethi, Mohammed
Sheikh-Jabbari, Charles Thorn and Herman Verlinde for helpful
discussions and comments. The work of SM is supported in part by DOE
grant DE-FG02-90ER40542 and by the Monell Foundation. The work of MR
and EV is supported by NSF-grant PHY-9802484 and DOE grant DE-FG02-91ER40571
respectively.

\end{document}